\documentclass[aip, amsmath, amssymb, reprint]{revtex4-1}

\usepackage{graphicx}
\usepackage{dcolumn}
\usepackage{bm}
\usepackage[utf8]{inputenc}
\usepackage[T1]{fontenc}
\usepackage{mathptmx}
\usepackage{etoolbox}
\usepackage{xcolor}

\usepackage{soul} 

\makeatletter
\def\@email#1#2{%
 \endgroup
 \patchcmd{\titleblock@produce}
  {\frontmatter@RRAPformat}
  {\frontmatter@RRAPformat{\produce@RRAP{*#1\href{mailto:#2}{#2}}}\frontmatter@RRAPformat}
  {}{}
}%
\makeatother

\begin{document}

\title{Quantifying ground state degeneracy in planar artificial spin ices: the magnetic structure factor approach}

\author{F. S. Nascimento}
\affiliation{Centro de Forma\c{c}ao de Professores, Universidade Federal do Rec\^{o}ncavo da Bahia, Amargosa, 45300-000, Bahia, Brazil}
\email{fabionascimento@ufrb.edu.br}

\author{L. B. de Oliveira}
\affiliation{Laborat\'{o}rio de Spintr\^{o}nica e Nanomagnetismo, Departamento de F\'{i}sica, Universidade Federal de Vi\c{c}osa, Vi\c{c}osa,	36570-900, Minas Gerais, Brazil}

\author{D. G. Duarte}
\affiliation{Laborat\'{o}rio de Spintr\^{o}nica e Nanomagnetismo, Departamento de F\'{i}sica, Universidade Federal de Vi\c{c}osa, Vi\c{c}osa, 36570-900, Minas Gerais, Brazil}

\author{C. I. L. de Araujo}
\affiliation{Laborat\'{o}rio de Spintr\^{o}nica e Nanomagnetismo, Departamento de F\'{i}sica, Universidade Federal de Vi\c{c}osa, Vi\c{c}osa,	36570-900, Minas Gerais, Brazil}

\author{W.A. Moura-Melo}
\affiliation{Laborat\'{o}rio de Spintr\^{o}nica e Nanomagnetismo, Departamento de F\'{i}sica, Universidade Federal de Vi\c{c}osa, Vi\c{c}osa,	36570-900, Minas Gerais, Brazil}

\author{A. R. Pereira}
\affiliation{Laborat\'{o}rio de Spintr\^{o}nica e Nanomagnetismo, Departamento de F\'{i}sica, Universidade Federal de Vi\c{c}osa, Vi\c{c}osa, 36570-900, Minas Gerais, Brazil}
\email{apereira@ufv.br}


\date{\today}

\begin{abstract}
Magnetic structure factor (MSF) is employed to investigate the ground state degeneracy in rectangular-like artificial spin ices. Our analysis considers the importance of nanoislands size via dumbbell model approximation. Pinch points in MSF and residual entropy are found for rectangular lattices with disconnected nanoislands, signalizing an emergent gauge field, through which magnetic monopoles interact effectively. Dipole-dipole interaction is also used and its predictions are compared with those obtained by dumbbell model.
\end{abstract}

\maketitle

\section{\label{sec:level1}Introduction}
Spin ice systems are magnetic structures in which the magnetic moments exhibit a complex behavior due to geometric frustration, which introduces additional degrees of freedom into the system \cite{harris_geometrical_1997, bramwell_spin_2001}. Typically, these systems display a substantial level of macroscopically degenerate ground state, also referred to as zero-point entropy \cite{pauling_structure_1935, ramirez_zero-point_1999, andrews_monte_2009}. In addition to this residual entropy, frustration can lead to other intricate behaviors, such as an emergent gauge field with generation of magnetic excitations that interact through a Coulomb potential \cite{castelnovo_magnetic_2008,henley_coulomb_2010}, the establishment of an effective ferromagnetic coupling  \cite{macedo_apparent_2018}, and the emergence of spin waves with well-defined frequencies \cite{li_writable_2022}. These spin waves can potentially offer versatile and programmable nanostructured waveguides and they may be important in specific magnonics mechanisms, among other functionalities. Hence, it is crucial to understand the physical properties of systems in which geometrical frustration takes place.

With the advancement of nanolithography techniques, the study of spin ice has been extended to the so-called artificial spin ices (ASI's) \cite{wang_artificial_2006,skjaervo_advances_2020,nisoli_colloquium_2013}. This attention is attributed to the relative ease with which these artificial systems can be fabricated in various geometries and to the capability of directly mapping their configurations in real space \cite{wang_artificial_2006, mengotti_real-space_2011} and in real time \cite{farhan_exploring_2013}. An ASI is composed by magnetic nanoislands (generally, made from ferromagnetic material, e.g., permalloy) arranged according to a certain underlying geometry. For instance, square ASI is the simplest planar geometry that incorporates features similar to those of water ice, such as having a ground state that follows the ice rule. However, the frustration in square ASI is not intense enough to produce macroscopically degenerate ground states \cite{wang_artificial_2006, moller_artificial_2006, budrikis_diversity_2011}. As a result, excitations in square spin ice are confined like a Nambu magnetic monopole pair by strong string tension \cite{mol_magnetic_2009, silva_thermodynamics_2012,silva_nambu_2013,morgan_thermal_2010,morley_thermally_2019,nambu_strings_1974}. Recently, advancements in experimental techniques have allowed the construction of three-dimensional artificial spin ice structures. In this context, two sublattices of nanomagnets are vertically separated by a small distance (a height offset $h$). Theoretical calculations \cite{moller_artificial_2006, mol_conditions_2010} and experimental results \cite{ perrin_extensive_2016, farhan_emergent_2019} have revealed evidences of free magnetic monopoles for a critical value of $h$. Indeed, experiments have shown unambiguous signatures of a Coulomb phase and algebraic spin-spin correlations, which are characterized by the presence of pinch points in the magnetic structure factor (MSF), in contrast to what occurs in the two-dimensional square ASI. So, considering again the planar case, rectangular spin ices may also bear ground state degeneracy \cite{nascimento_confinement_2012, ribeiro_realization_2017}. Specifically, when the ratio of horizontal to vertical lattice constants reaches the value of $\sqrt3$, a point-dipole model indicates a degeneracy of vertex types obeying the ice rule.
\begin{figure*}
\includegraphics[width=420 pt]{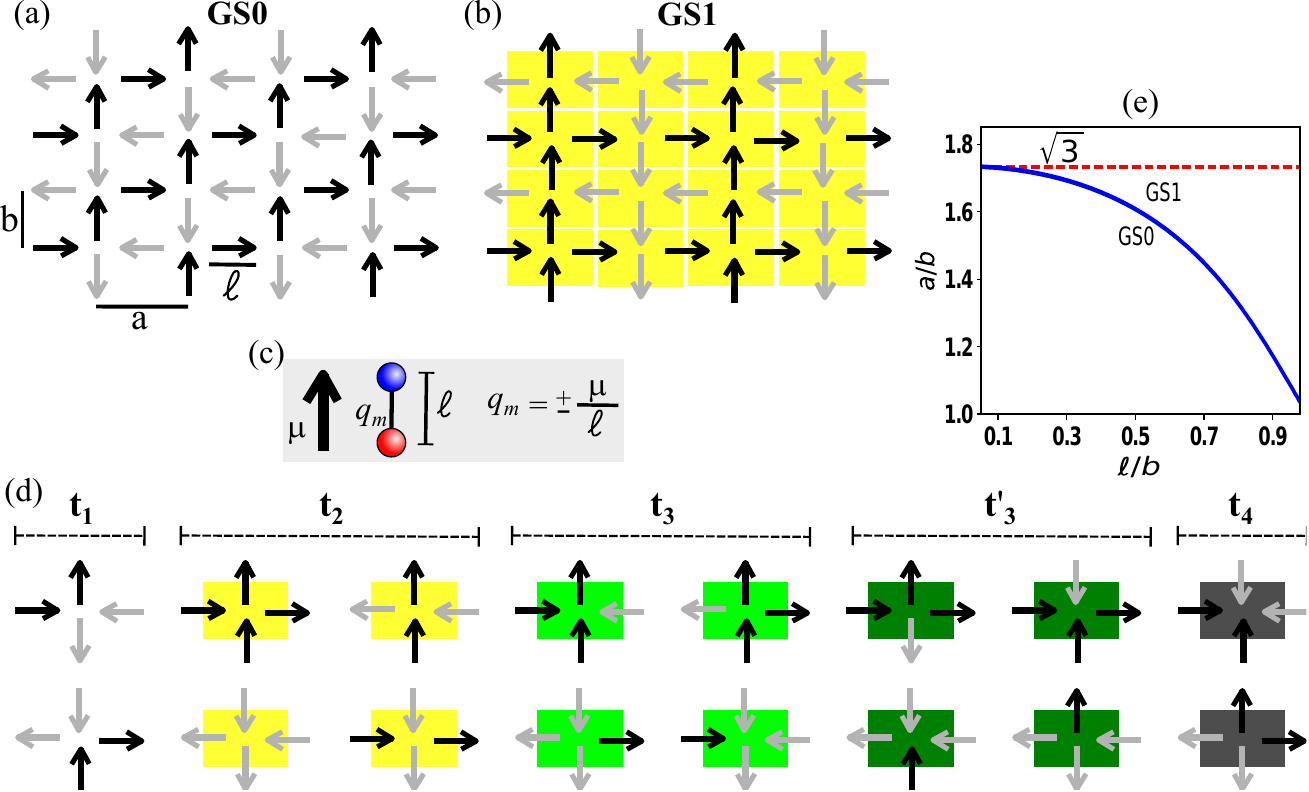}
\caption{\label{fig:fig1} "(a)-(b) The observed ground state configurations for the rectangular spin ice. (c) Each dipole can be represented by a pair of opposite magnetic charges (dumbbell model). (d) Shows the four possible vertex types in a single artificial spin ice (ASI) (derived from Hamiltonian (1)). In this case, the ground state is realized when all vertices are in configurations of class $t1$, adhering to the ice rule. Class $t2$ also includes vertices that follow the ice rule, but they possess higher energy compared to vertices belonging to class $t1$. Generally, vertices of type $t2$ are associated with strings connecting monopoles within the same ASI. Classes $t3$ and $t4$ violate the ice rule and contain excited states (magnetic monopoles). (e) Favorable ground state as function of geometric lattice ratio, $\gamma=a/b$, and island size to separation ratio, $\ell/b$. The relative energy, $(E_{\rm GS1}-E_{\rm GS0})/E_{\rm GS1}$, is plotted in the $\gamma \times \ell/b$ plane. Along the blue critical line, both configurations share the same energy. Below such a curve, \textbf{GS0} is energetically favorable, whereas above \textbf{GS1} comes to be the true ground state.}
\end{figure*}

Once artificial systems can be directly visualized in real space, their associated MSF's can be eventually constructed in the usual way \cite{farhan_emergent_2019,perrin_extensive_2016,rougemaille_magnetic_2021}. This capability provides a powerful tool for exploring pairwise spin correlations in reciprocal space ($k$-space). For example, in cases where the ground state is ordered, MSF is composed of distinct magnetic Bragg peaks. However, when the ground state takes on a disordered nature, resembling a spin liquid, MSF takes on a diffuse quality while still manifesting structural attributes. Here, we shall verify these characteristics in the outcomes obtained from the analysis of artificial rectangular-type ASI's.

\section{Model and Methods}

A rectangular-like spin ice \cite{nascimento_confinement_2012, li_geometry_2011, ribeiro_realization_2017} is characterized by a horizontal lattice constant $a$ and a vertical lattice constant $b$ (Figs 1(a) and 1(b)). The square lattice is readily obtained by setting $a=b$. In our simulations, parameter $b$ is fixed whereas $a$ is varied in the interval $(b, 2b)$. The nanoisland length is specified by a value $\ell$ (Fig 1(a)) within the interval $(0, b)$. We consider a lattice comprising $20\times20$ vertices and 840 spins. Note that, if $\ell=a=b$, the nanoislands are physically connected and this system is in sharp contrast with the usual cases \cite{perrin_quasidegenerate_2019}. Since we are interested in possible formation of magnetic monopoles in the lattice vertices, disconnected nanoislands should be more appropriated objects for investigation. Here, each spin (i.e., the magnetic moment of a nanoisland) interacts with all the others in the lattice. Both, point-dipole and the dumbbell models are used to describe spin interactions, and open boundary conditions are assumed. The dipole-dipole interaction is given by:
\[ \displaystyle H=\frac{\mu_0}{4\pi}\sum_{ij}\Big[\frac{\vec\mu_i\cdot\vec\mu_j}{r^3_{ij}} - \frac{3(\vec\mu_i\cdot\vec r_{ij})(\vec\mu_j\cdot\vec r_{ij})}{r^5_{ij}} \Big]\,, \]
where $\vec\mu_i$ is the magnetic moment at site $i$ and $\vec r_{ij}$ is a vector associated to the distance between two sites $i$ and $j$.  In the dumbbell model framework\cite{castelnovo_magnetic_2008, moller_magnetic_2009}, a dipole is represented by a pair of spaced opposite charges $q_m=\pm\mu/\ell$ (Fig. 1(c)). Whenever $\ell \ll 1$, the dumbbell model reproduces the (point-like) dipole-dipole interaction. In turn, charges within dumbbell description interact according to:
\[ \displaystyle H= \frac{\mu_0}{4 \pi}\sum_{ij}\frac{q_iq_j}{r_{ij}} \]

\begin{figure*}
\includegraphics[width=0.95\textwidth]{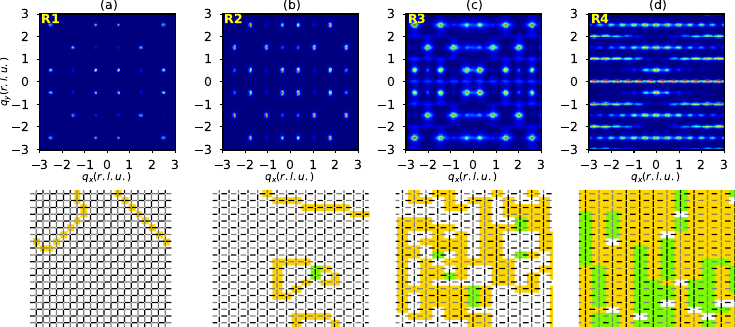}
\caption{\label{fig:fig2} (upper panels) depict MSF for the 4 arrangments, $R_1$, $R_2$, $R_3$, and $R_4$. Namely, notice the drastic contrast in the MSF patterns as the arrangement is stretched. (lower panels) display typical collective spin configuration for each of these arrangements. MSF has been computed using a $120\times120$ points matrix covering an area of $q_x, q_y$ in the interval $[-6\pi, 6\pi]$.}
\end{figure*}

Figure \ref{fig:fig1}(d) illustrates all possible vertex types for the planar rectangular spin ice. A total of 5 vertex types can be realized: the first two, $t_1$ and $t_2$, adhere to the ice rule ($2in-2out$), while the following two, $t_3$ and $t'_3$, represent excitation showing up as effective magnetic poles carrying single charge, whereas $t_4$ is regarded with the appearance of double-charged excitation. Such vertices are expected to take place only at high temperatures, once they are the most energetic ones. The energy of these vertex types depends on the geometric factor \cite{nascimento_confinement_2012} $\gamma=a/b$. For $\gamma$ smaller than a critical value $\gamma_c$, the ground state is populated by $t_1$ vertices, resulting in the \textbf{GS0} configuration (Fig. 1(a)). However, when $\gamma > \gamma_c$, $t_2$ becomes energetically more favorable than $t_1$. This leads to a ground state transition from \textbf{GS0} to \textbf{GS1} configuration, Fig. 1(b). To better understand the scenario, we initially note that for $\gamma < \gamma_c$, the ground state is the same as that of the square spin ice (\textbf{GS0}), which is doubly degenerate. On the other hand, for $\gamma > \gamma_c$, the ground state exhibits \textbf{GS1} pattern with ferromagnetic stripes, resulting in a quadruple degeneracy. Point-like dipole-dipole interaction predicts that the energies of \textbf{GS0} and \textbf{GS1}, $E_{GS0}$ and $E_{GS1}$ respectively, are equal at $\gamma_c=\sqrt{3}$. Consequently, the coexistence of both states yields degenerate ground state. Although no residual entropy has been predicted for this case, we shall provide further evidence for such a degeneracy in MSF. As will be discussed later, the scenario becomes more interesting whenever nanoislands size $\ell$ is taken into account. 

The magnetic structure factor is defined like follows:
$$I(\vec q) = \frac1N \sum_{i=1}^N\sum_{j=1}^N \vec{S}_i^{\perp}\cdot \vec{S}_j^{\perp} \exp({i\vec q\cdot\vec r_{ij}})\,,$$
where $\vec S_i^{\perp} = \vec S_i - (\hat q\cdot \vec S_i)\hat q$ represents the component of the spin vector of each island, $\vec S_i$, perpendicular to the reciprocal space vector $\vec q$. The unit vector is given by $\hat q= \vec q/||\vec q||$; $\vec r_{ij}$ is the vector from island at site $i$ to to that at $j$; $N$ is the total number of islands (spins). The main advantage of MSF approach relies on the fact that its results may be directly compared to cross-section measurements, for instance, by means of neutron scattering, as recently performed in Ref. \cite{farhan_emergent_2019} for ASI's systems.

Firstly, one obtains low-energy spin configurations as a function of $\gamma$. This is accomplished using a standard Monte Carlo technique along with the Metropolis algorithm, implemented using the Boltzmann distribution $\sim e^{-\Delta E/k_B T}$. The process begins with a disordered state at a high temperature, and then the system undergoes slow dynamics by gradually cooling it to a very low temperature, $ \sim 0.1D/k_B$. Later, MSF is evaluated for each rectangular ASI (an average over 100 identical samples is employed).

\section{Results and discussion}
\subsection{Point-like dipole model}
We start by considering the point-like dipole model and analyze MSF as a function of lattice aspect ratio $\gamma$. In Fig. 2, upper panels, MSF is depicted for $R_1$, $R_2$, $R_3$ e $R_4$ arrangements ($R_1$ for $\gamma=a/b=\sqrt1$, $R_2$ for $\gamma=a/b=\sqrt2$, and so forth). In the lower panels, spin configurations achieved after annealing are shown for each of these $\gamma$. The low-energy configuration of $R_1$ and $R_2$ arrangements are mostly populated by $t_1$ vertices, while $t_2$ is most abundant in $R_4$. [Some vertices are in their excited state due to the low-energy regime used to obtain the simulated samples]. The first excited state requires the presence of $t_2$ and $t_3$ vertices: $t_3$ supports a monopole-antimonopole pair (with unity and opposite charges $\pm 1$, indicated by green spots) connected by an energetic string, which is a segment of $t_2$ vertices (yellow spots). MSF for square ASI, $R_1$, is characterized by magnetic Bragg peaks at the corners of the Brillouin zone, evidencing an ordered state whose monopoles are bound in pair joined by strong string tension. For $R_2$ a similar pattern is realized, but with narrower Bragg peaks along the $q_x$ axis, suggesting that string tension is diminishing, as expected for a rectangular geometry. This is prominently observed in $R_3$, where broaden spots weakly connected take the place of sharply peaks. This is commonly associated to a degenerate ground state. Average MSF for R4 consists of random arrangements of fully polarized lines, and the associated MSF exhibits well-defined lines. Notably, magnetic Bragg peaks are absent in this MSF due to the absence of antiferromagnetic long-range order.

\subsection{Dumbbell model}
Now, we would like to understand how the dipoles length modifies the energetic of the whole ASI system. Indeed, energy evaluation within dumbbell model clearly depends upon dipole length, $\ell$. Thus, in order to correctly evaluate the energy of a certain arrangement, we need to specify both, the geometric lattice ratio, $\gamma=a/b$, and $\ell$. Figure 1(e) depicts the relative energy, $(E_{\rm GS1}-E_{\rm GS0})/E_{\rm GS1}$, as function of  $\gamma=a/b$ and $\ell/b$. Along the blue critical line, both \textbf{GS0} and \textbf{GS1}, share the same energy and they are expected to equally populate the ground state. 
State \textbf{GS0} (\textbf{GS1}) is energetically favorable below (above) such a curve, as indicated in the `phase diagram'. In our analysis, we have verified that whenever $\ell/b$ is small, energetic predicted by point-like dipole and dumbbell models agree very well. For instance, one obtains the critical geometric ratio $\gamma_c\approx\sqrt3$ for every $\ell<0.2b$.

\begin{figure}
\includegraphics[width=225 pt]{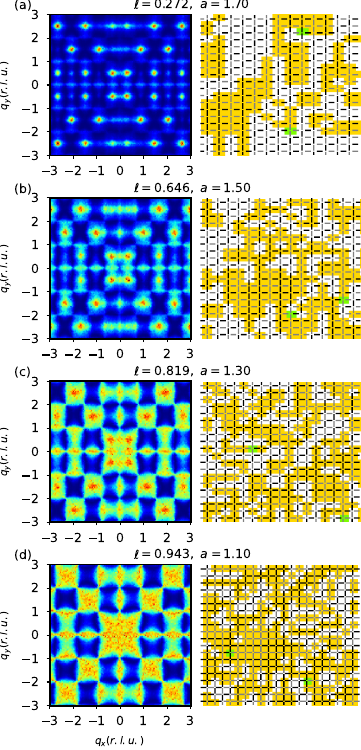}
\caption{\label{fig:fig3} MSF of low-energy states along the blue line from Fig. 3. Four different values of $(\ell/b,\, a/b)$ are taken: (a) $(0.272, \, 1.7)$,  (b) $(0.646,\,1.5)$, (c) $(0.819,\, 1.3)$, and (d) $(0.943,\, 1.1)$. [MSF is evaluated over a $120\times120$ matrix covering an area of $q_x, q_y$ in the interval $(-6\pi, 6\pi)$].}
\end{figure}

Figure 3 (a)-(d) shows how the spin configuration profile is modified with respect to the lattice geometry and dipole length. This fact is illustrated by considering four different states along the critical line and their reciprocal state signature as captured by MSF. For example, Fig. 3(a) shows a MSF pattern quite similar to that obtained for $\sqrt 3$ within the point-like dipole framework. However, these states become more densely populated as $\ell$ increases. For $\ell\ge 0.646b$, regions characterized by a pronounced narrowing, referred to as pinch points, start to be observed. The presence of these pinch points is a signature of an emergent gauge field (a Coulomb phase), implying that the planar rectangular spin ice system supports free magnetic charges. Indeed, the spin configurations in this regime have shown the presence of widely separated $t3$ vertices, which can only occur if the string tension vanishes. Such a configuration does not appear in square spin ice or even in rectangular spin ice (simulated with the point-like dipole model).

\begin{figure}
\includegraphics[width=250 pt]{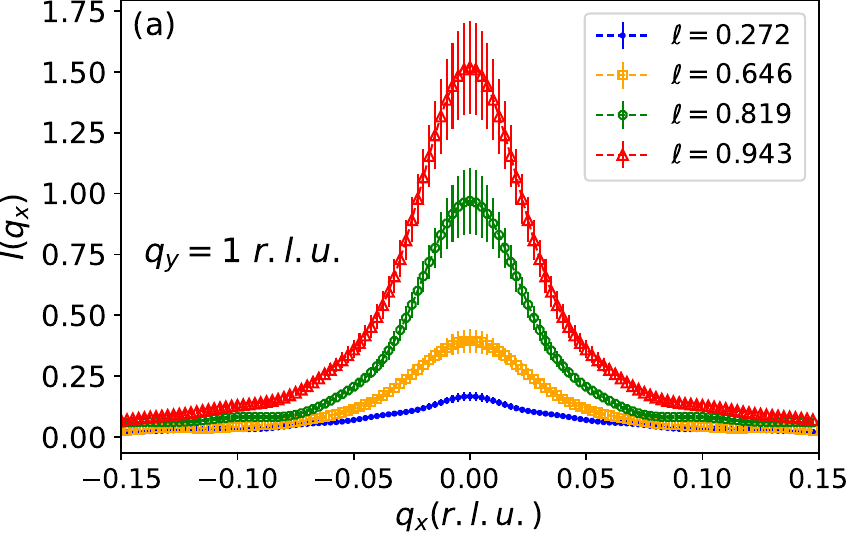}
\includegraphics[width=250 pt]{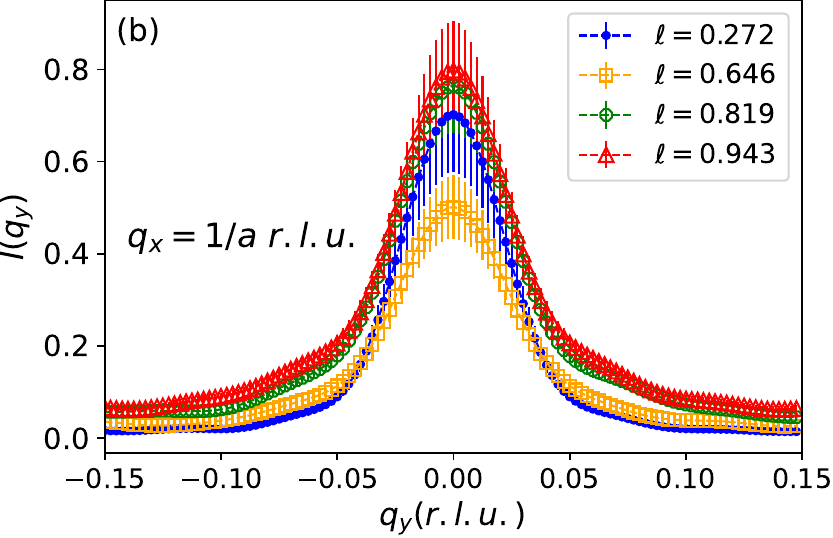}
\caption{\label{fig:fig4} Variation of MSF intensity: (a) along the top edge of the first Brillouin zone and (b) along the left edge of the first Brillouin zone. The error bars have been evaluated using the standard error from a set of 100 samples.}
\end{figure}

Besides changing the MSF pattern according to the dipole length, one realizes a significant increase in the MSF intensity in the pinch point region. Figure 4 shows the MSF intensities of the pinch points located on the top edge and the left edge of the first Brillouin zone. As the length of the dipole increases, we observe a substantial rise in intensity (see Fig. 4(a)). This effect is not as significant at the pinch point on the left edge, as shown in Fig. 4(b). For $\ell=0.272b$, it is worth noting that the pinch point on the side edge has an intensity peak of $I(q_y=0)=0.7(1)$, which is over four times bigger than the peak intensity of the pinch point on the top edge, which is $0.17(2)$. This observation corroborates previous findings \cite{nascimento_confinement_2012}, which suggest that monopoles became deconfined when separated vertically but remained somewhat confined when separated horizontally at R3 rectangular ASI when point-like dipole model describes interactions between spins. This confinement along the horizontal direction implies a small yet discernible string tension \cite{nascimento_confinement_2012}. However, as $\ell$ increases, the dumbbell model predicts that the monopoles tend to undergo deconfinement in both vertical and horizontal directions.

\begin{figure}
\includegraphics[width=240 pt]{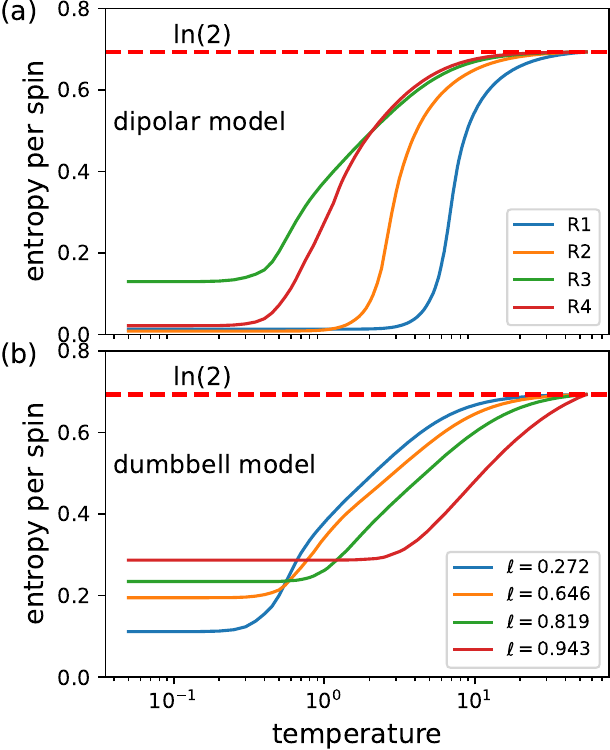}
\caption{\label{fig:fig5} Entropy per spin as a function of the temperature for (a) dipole model and (b) dumbbell model. The entropy per spin was found by integrating the specific heat over temperature ($c_V/T$), in which the temperature varies from $0.1$ to $50~D/k_B$. We have imposed the condition that, at high temperatures, the entropy per spin is equal to $s = \ln(2)$. Thus, we have determined the residual entropy per spin for dipolar model $s_{0,R1}=0.013$, $s_{0,R2}=0.009$, $s_{0,R3}=0.130$ e $s_{0,R4}=0.021$. For dumbbell model, we found $s_0=0.112$ for $\ell=0.272~b$, $s_0=0.195$ for $\ell=0.646~b$, $s_0=0.234$ for $\ell=0.819~b$ and $s_0=0.287$ for $\ell=0.943~b$}
\end{figure}

Figure 5 shows the entropy $s$ per spin as a function of temperature for both the point-like dipole model and the dumbbell model. In our numerical calculations, we utilize the high-temperature paramagnetic phase to impose constraints on entropy. In this phase, each spin orientation remains independent. Consequently, at high temperatures, the entropy per spin is given by $s = \ln 2$. Thus, we can measure the residual entropy $s_0$ at low temperatures by integrating the specific heat over temperature ($\Delta S = \int_{T_i}^{T_f}(c_V/T)dT$). In the point-like dipole model (see Fig. 5(a)) we notice that the residual entropy reaches its maximum value at the R3 geometry, where we obtain approximately $s_0\approx 0.13~units$. However, for R1, R2, and R4, we observe that the residual entropy is approximately zero. As expected, these cases do not exhibit a highly degenerate ground state.  For dumbbell model, we have measured the entropy for the same four dipole lengths analyzed previously. We notice that the residual entropy increases as the dipole length becomes larger. Specifically, for $\ell=0.943b$, the estimated residual entropy is approximately $s_0\approx 0.29~units$. It is more than two times larger than that observed for the R3 geometry for point-like dipole system.

\section{Conclusion}
In summary, we have compared the magnetic structure factor obtained from (point-like) dipole-dipole and dumbbell models. Using pinch point formation predicted by the dumbbell model, we can emphasize the parameters of the artificial rectangular spin ice system that maximizes the ground state degenerancy. Similar results have been observed recently in square three-dimensional artificial spin ice\cite{perrin_extensive_2016}. We have also calculated the residual entropy of these systems. In general, our results indicated that an emergent Coulomb phase is also possible in planar artificial spin ices with disconnected nanoislands, displayed in a  rectangular lattice, which provides a simpler structure to be experimentally constructed. So, the ground state is degenerate in such a way that the string tension vanishes, and free magnetic monopoles may appear in the vertices of the rectangular lattices. As the nanoislands size $\ell$ increases, pinch points and residual entropy are obtained for smaller ratios $a/b$ (for instance, when $\ell /b = 0.943$, $a/b = 1.1$). So, in the limit $\ell/b=a/b=1$, our results approach the situation of a square lattice with physically connected nanomagnets \cite{perrin_quasidegenerate_2019}. To conclude, a phenomenological description of MSF of rectangular spin ice in the scenario of large nanoislands, is presented, providing an empirical understanding of the emerging spin-spin correlations within the spin ice manifold.

\begin{acknowledgments}
The authors would like to thank Brazilian agencies CNPq, FAPEMIG, and CAPES, and INCT (Spintronics and advanced nanomagnetic materials) for partial financial support.
\end{acknowledgments}


%

\end{document}